\documentclass[twocolumn,aps,superscriptaddress]{revtex4-1}
\usepackage{palatino}
\usepackage[latin1]{inputenc}
\usepackage{epsf}
\usepackage{amsmath,amssymb}
\usepackage{latexsym}
\usepackage{calc}
\usepackage{color}
\usepackage{shadow}
\usepackage{epsfig}

\newcommand{\ben}{\begin{equation}}
\newcommand{\een}{\end{equation}}
\newcommand{\bea}{\begin{eqnarray}}
\newcommand{\eea}{\end{eqnarray}}

\def\sss{\scriptscriptstyle\rm}

\def\c{_{\sss C}}
\def\s{_{\sss S}}
\def\xc{_{\sss XC}}
\def\Hx{_{\sss HX}}
\def\Hxc{_{\sss HXC}}

\def\br{{\bf r}}

\begin{document}

\title{Charge-Transfer in Time-Dependent Density Functional Theory: Insights from the Asymmetric Hubbard Dimer}
\author{J. I. Fuks}
\author{N. T. Maitra}
\affiliation{Department of Physics and Astronomy, Hunter College and the City University of New York, 695 Park Avenue, New York, New York 10065, USA}
\date{\today}

\begin{abstract}
We show that propagation with the best possible adiabatic
approximation in time-dependent density functional theory fails to
properly transfer charge in an asymmetric two-site Hubbard model. The
approximation is adiabatic but exact otherwise, constructed from the
exact ground-state exchange-correlation functional that we compute via
constrained search.  The model shares the essential features of
charge-transfer dynamics in a real-space long-range molecule, so the
results imply that the best possible adiabatic approximation, despite
capturing non-local step features relevant to dissociation and
charge-transfer excitations, cannot capture fully time-resolved
charge-transfer dynamics.
\end{abstract}
\maketitle

Charge-transfer (CT) dynamics are increasingly important in biology,
chemistry and physics, underlying critical processes in photovoltaics,
vision, photosynthesis, molecular electronics, and the control of
coupled electron-ion dynamics
(e.g. Refs~\cite{DP07,Jailaubekov12,Polli,NR03,Sansone10}).  Yet an
accurate theoretical description, capturing correlated electron
motion, is notoriously difficult especially over large distances. For
most applications, a time-resolved picture is crucial, and the systems
are large enough that time-dependent density functional theory (TDDFT)
is the only calculationally feasible
approach~\cite{RG84,TDDFTbook,Carstenbook}. Standard functional
approximations underestimate CT excitations, but improved functionals
have been developed~\cite{SKB09p}. Still, a truly time-resolved
description must go beyond a calculation of the excitation spectrum:
electron transfer between regions of space is clearly
non-perturbative.  TDDFT certainly applies in the non-linear regime,
and has given useful predictions in many cases, including CT
dynamics~\cite{Rozzi13}.  At the same time, there is a dearth of
alternative accurate practical methods to test TDDFT
calculations. Results on simplified exactly-solvable model systems are
not always optimistic~\cite{RB09,FHTR11,EFRM12,FERM13}.

Almost all non-perturbative TDDFT calculations utilize an adiabatic
 exchange-correlation potential: $v\xc^{\rm
  A}[n;\Psi_0,\Phi_0](\br,t) = v\xc^{\rm gs}[n(t)]({\bf r})$.
Errors arise  from two distinct sources: one
is the choice of the ground-state (gs) functional approximation,
the other is the adiabatic approximation itself. To separate
these the {\it adiabatically-exact} (AE)
approximation~\cite{TGK08} is defined: the instantaneous density is input into
the exact gs functional,
$v\xc^{\rm AE}[n;\Psi_0,\Phi_0](\br,t)=v\xc^{\rm AE}[n](\br,t) =
v\xc^{\rm exact\;gs}[n(t)]({\bf r})$.  This approximation neglects
memory-effects (dependence on the density's history and true and Kohn-Sham (KS) initial states $\Psi_0$ and $\Phi_0$) but is fully non-local in space.
%and, if the true and KS states at time $t$ were actually gs's of some potential,it would be exact at time $t$.  
Finding $v\xc^{\rm AE}[n](\br,t)$ requires an iterative density-inversion scheme to find interacting and non-interacting gs's of a given
density, and it has been done just for a few model
systems~\cite{TGK08,EFRM12,FERM13}. Usually one evaluates the
AE potential on the exact density $n(t)$, 
$v\xc^{\rm AE}[n(t)](\br)$, and compares with the exact potential
$v\xc[n,\Psi_0,\Phi_0](\br,t)$ at that time to analyse how good the
approximation is.
A more useful
assessment would be to self-consistently propagate the
KS orbitals with it, using at each time-step, the AE potential
evaluated on the self-consistent instantaneous density.  This clearly
requires much more numerical effort, as many iterations need to be
performed at every time-step to find the potential to propagate in;
it has only been done few examples ~\cite{TGK08, TK09b, RP10}.  For CT it is particularly challenging to converge the
iterations, due to the very low density between the atoms.

For a model molecule composed of
closed-shell atoms and driven at the CT resonance, a step associated
with the CT process gradually builds up over time in the exact
correlation potential~\cite{FERM13}.  The AE approximation fails to capture the
dynamical step of Refs.~\cite{EFRM12,LFSEM13} but, when evaluated on
the {\it exact} density, does show a CT step, although of a smaller
size than the exact.  Available approximations do not yield any step
structure whatsover, and the dismal failure of ALDA, ASIC-LDA, and
AEXX, to transfer any charge was shown in Ref.~\cite{FERM13}
and attributed to this lack of step structure.  We expect some blame
must go to the adiabatic approximation itself, but is the partial step
of the AE approximation enough to give a reasonable description of the
CT dynamics? If yes, this would greatly simplify the on-going search
for accurate functionals for non-perturbative CT.  To answer the
question, we must propagate with the AE self-consistently, but as
discussed above, this procedure is numerically very challenging for CT
dynamics.  We show here that the answer is no, by studying
CT in a two-fermion asymmetric Hubbard dimer, which shares
the essential features of CT dynamics in real-space molecules. Due
to the small Hilbert space of the dimer the exact gs functional can be
found and used in $v\xc^{\rm AE}(t)$ to self-consistently propagate
the system. We can then assess errors in the adiabatic approximation
for CT dynamics independently of those due to the gs approximation
used. We find the adiabatic approximation is inherently poor, and
analyze the potentials to explain why.

The Hamiltonian of the
two-site interacting Hubbard model with on-site repulsion $U$ and hopping 
parameter $T$ \cite{AG02,LU08,V08,CF12,FT12,FFTAKR13,B08,CC13} is:
\begin{align}
 \hat{H}= & -T \sum_\sigma \left( \hat{c}_{L\sigma}^\dag \hat{c}_{R\sigma} +\hat{c}_{R\sigma}^\dag\hat{c}_{L\sigma} \right)
+ U \left( \hat{n}_{L \uparrow} \hat{n}_{L \downarrow} + \hat{n}_{R \uparrow} \hat{n}_{R \downarrow}\right)  \nonumber \\ 
          & +  \frac{\Delta v (t)}{2}(\hat{n}_{L} -\hat{n}_{R}) ,
%& + \left( \frac{\Delta v^{0}}{2} + {\mathcal E(t)}\right)(\hat{n}_{L} -\hat{n}_{R}) ,
\label{eq:HubbardH}
\end{align}
where $\hat{c}_{L(R)\sigma}^{\dag}$ and $\hat{c}_{L(R)\sigma}$ are creation and 
annihilation operators for a spin-$\sigma$ electron on the left(right) site $L(R)$, respectively, and $\hat{n}_{L(R)}=\sum_{\sigma=\uparrow, \downarrow}\hat{c}_{L(R)\sigma}^{\dag} \hat{c}_{L(R)\sigma}$ are the site-occupancy operators.
 The dipole $\langle \hat{n}_{L} -\hat{n}_{R}\rangle = \Delta n$ is the main variable~\cite{FFTAKR13}; the total number of fermions is fixed at 2. A static potential between the  sites, $\Delta v^{0}= \sum_\sigma (v^0_{L\sigma} - v^0_{R \sigma})$, renders the Hubbard dimer asymmetric. The external potential $\Delta v (t)$ is given by $\Delta v(t)= \Delta v^{0}+ 2{\mathcal E(t)}$.
The long-range molecule is modeled by $T/U \to 0$: for fixed  $U$,  $T \to 0$ corresponds to a large separation between the sites (equivalent to the strongly correlated limit $U \to \infty$). 
We choose $T=0.05$, use $\hbar=e=1$ throughout, and energies are given in units of $U$.

\begin{figure}[h]
\begin{center}
\includegraphics[width=0.3\textwidth]{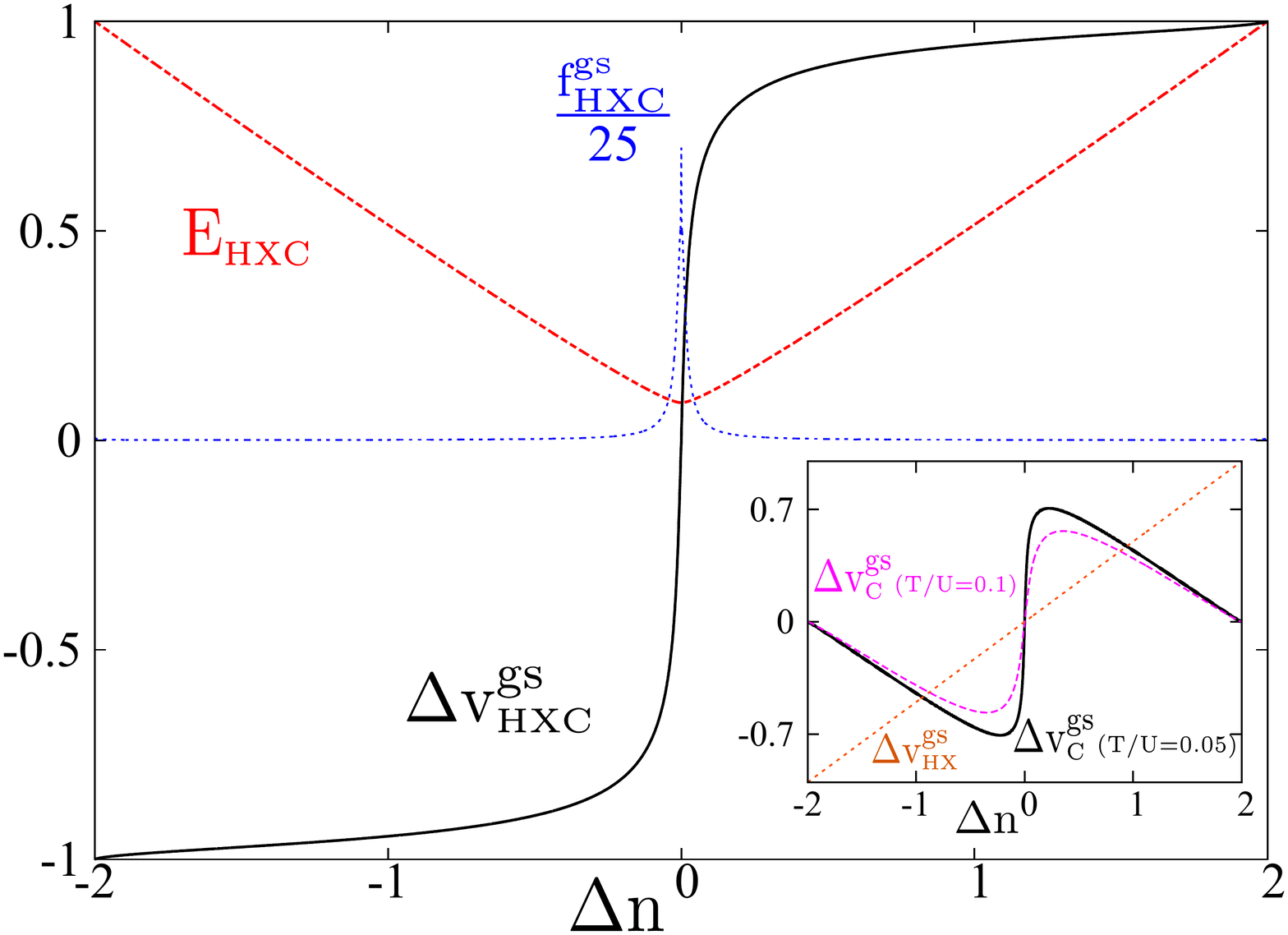}
\end{center}
\caption{The exact  $E\Hxc[\Delta n]$ (red dashed),  gs potential  $\Delta v\Hxc^{\rm gs}[\Delta n]$ (black solid) and scaled gs kernel $f^{\rm gs}\Hxc[\Delta n]/25$ (blue dotted) for  $T/U = 0.05$. The inset shows the correlation potential $\Delta v\c^{\rm gs}[\Delta n]$ for $T/U=0.05$ (black solid), $T/U=0.1$ (pink dashed), and the HX potential $\Delta v^{\rm gs}\Hx[\Delta n]$ (orange dotted). All functionals are in units of $U$.}
\label{fig:EHxc_vHxc_vc}
\end{figure}

The singlet sector of the vector space is three-dimensional, enabling an exhaustive search over all wavefunctions
to find the exact Hartree-exchange-correlation (HXC) energy
functional, $E\Hxc[\Delta n]$, plotted in Figure~\ref{fig:EHxc_vHxc_vc}. This  follows the procedure of Ref.~\cite{FFTAKR13}.

As $T/U$ decreases the energy becomes sharper (more V-like) at $\Delta n=0$ (Fig.~\ref{fig:EHxc_vHxc_vc}),
while the  potential $\Delta v\Hxc^{\rm gs}[\Delta n]$ approaches a step function there, contained in the correlation potential (see inset Figure ~\ref{fig:EHxc_vHxc_vc}). This indicates the derivative-discontinuity of the one-electron site, as will be discussed shortly. Note that $\Delta v\Hxc = \Delta v\Hx + \Delta v\c$ where $\Delta v\Hx = U\Delta n/2$~\cite{FFTAKR13}.

The KS Hamiltonian has the form of Eq.~(\ref{eq:HubbardH}) but with $U
= 0$ and $\Delta v(t)$ replaced by $\Delta v\s[\Delta n, \Phi_{\rm
    gs}](t) = v\Hxc[\Delta n, \Psi_{\rm gs}, \Phi_{\rm gs}](t) +
\Delta v(t)$, defined such that the interacting density $\Delta n(t)$
is reproduced. A self-developed code in second quantization, using a
Crank-Nicholson propagator and a 0.01 time step, was used for the
propagations.

To model closed-shell to closed-shell CT (cs--cs) in a real molecule,
we take $\Delta v^{0} = -2.0~U$ where the gs has $\Delta n_{\rm gs}
=1.9901$ and study the transition to the CT excited state with $\Delta
n_{\rm CT} = 0.0090$ and frequency $\omega_{\rm CT}=1.0083~U$; we take
$\mathcal{E}(t) = 0.2\sin(1.0083 U t)$.  For open-shell to
open-shell (os--os) CT in a real molecule, we instead take $\Delta
v^{0} = -0.4 ~U$ resulting in a slightly asymmetric gs $\Delta n_{\rm
  gs} = 0.02137$, and study the transition to the CT excited state
where $\Delta n_{CT}= 1.9734$ and $\omega_{\rm CT}=0.6199~U$; here we
take $\mathcal{E}(t) = 0.18\sin(0.6199 U t)$.  In either case
the field $\mathcal{E}(t)$ is resonant with magnitude weak enough such that
only the ground and above-mentioned CT states are significantly
occupied during the dynamics.

\begin{figure}[h]
\begin{center}
\includegraphics[width=0.5\textwidth, height=0.2\textwidth]{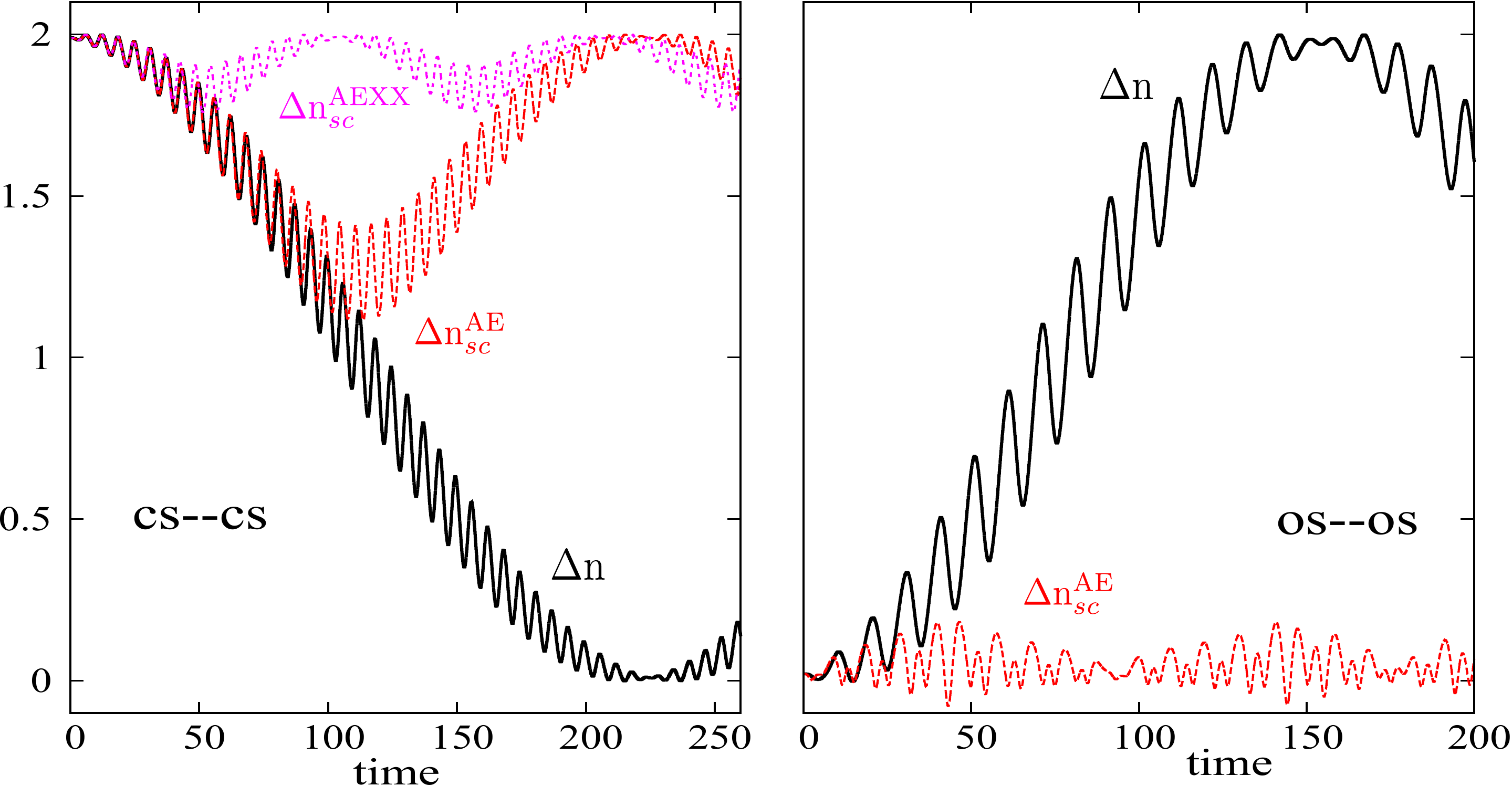}
\end{center}
\caption{ Exact dipole $\Delta n$ (black solid),  AE dipole, $\Delta n^{\rm AE}_{sc}$ (red dashed) and the AEXX dipole $\Delta n^{\rm AEXX}_{sc}$ (pink dotted). Left: cs--cs CT. Right: os--os CT.
Time is in units of $1/U$.}
\label{fig:dipoles}
\end{figure}

\underline{\it cs--cs CT} The dipoles are shown on
the left panel of Figure~\ref{fig:dipoles}; the CT excited
state is reached at around $t=224/U$.
The similarity of the exact dipole $\Delta n(t)$ with the real-space dynamics of Figure 4 of
Ref.~\cite{FERM13} is evident; also the
adiabatic exact-exchange (AEXX) dipole on the left of Fig.~\ref{fig:dipoles}
drastically fails to complete the CT, resembling the real-space AEXX
case. 
Propagating the KS system with the AE functional,
obtained at each time-step by inserting the instantaneous density  $\Delta n^{\rm AE}_{sc}$ into the exact gs HXC potential 
$\Delta v\Hxc^{\rm gs}[\Delta n^{\rm AE}_{sc}]$ of Fig.~\ref{fig:EHxc_vHxc_vc}, we obtain $\Delta n^{\rm AE}_{sc}$ on the left of Fig.~\ref{fig:dipoles}.  $\Delta n^{\rm AE}_{sc}$ follows the exact for a
longer time than the AEXX, but ultimately fails to complete the CT.  The AE propagation,
 shows that one must go beyond the
adiabatic approximation to correctly describe CT.

\begin{figure}[h]
\begin{center}
\includegraphics[width=0.5\textwidth,height=0.35\textwidth]{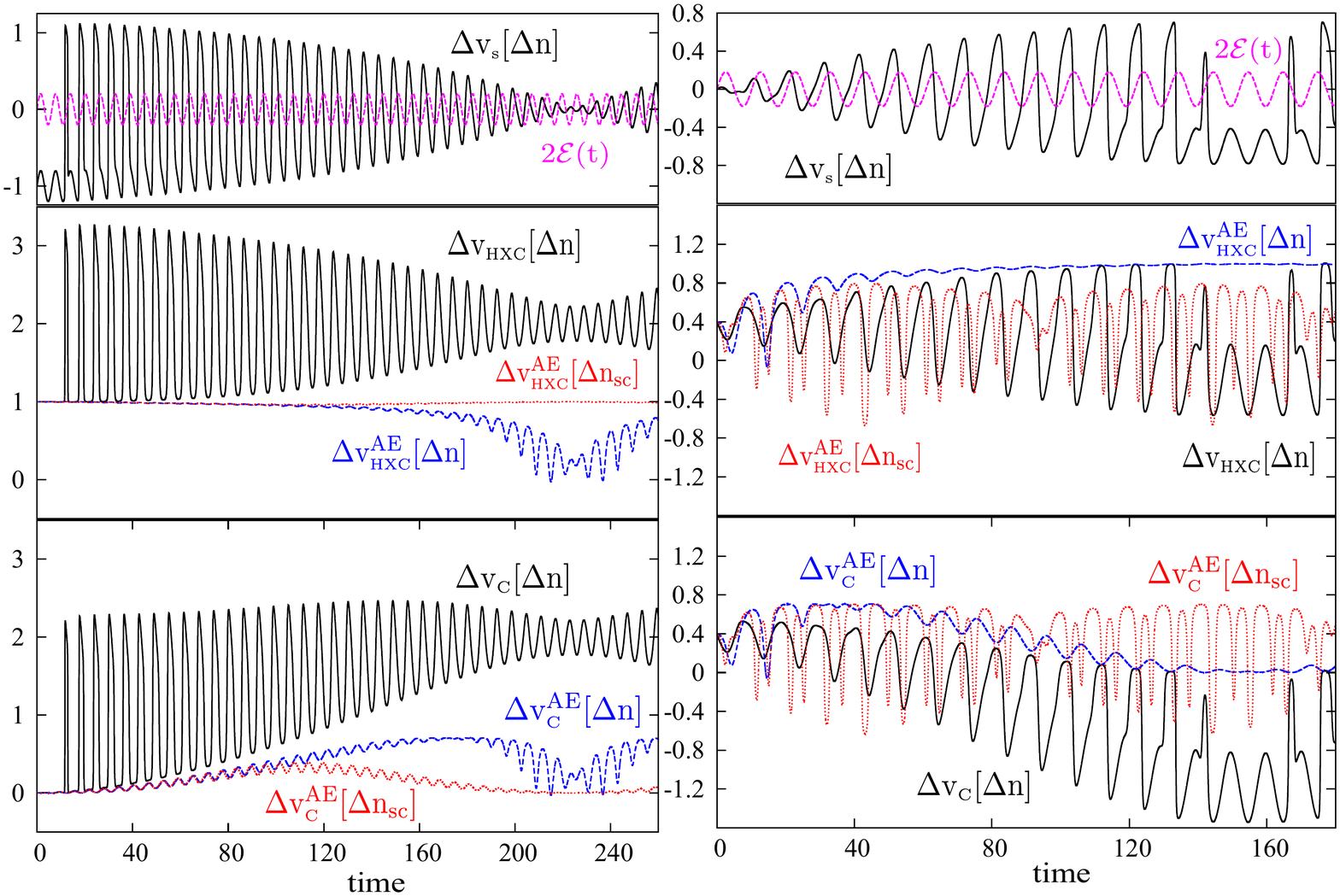}
\end{center}
\caption{
Upper panel: exact KS potential $\Delta v\s$ (black solid) (first term in RHS Eq.~(\ref{eq:vhxc})), and the  field $\mathcal{E}(t)$ (pink dashed).
Middle panel: exact HXC potential Eq.~(\ref{eq:vhxc}) (black solid),  the AE HXC potential  $\Delta v\Hxc^{\rm AE}[\Delta n]$ 
(blue dashed) and  AE HXC potential $\Delta v\Hxc^{\rm AE}[\Delta n_{sc}]$ (red dotted), with $\Delta n_{sc}=\Delta n_{sc}^{\rm AE}$. Lower panel: correlation potentials.
Left: cs--cs CT. 
Right:  os--os CT.}
\label{fig:vHxc_AE_AEsc}
\end{figure}

We plot the exact and AE potentials on the left of
Figure~\ref{fig:vHxc_AE_AEsc}. The top left panel shows the exact KS
potential alongside the applied field. The middle left shows
the exact HXC potential $\Delta v\Hxc[\Delta n](t)$, the AE
potential evaluated on the exact density $\Delta v\Hxc^{\rm AE}[\Delta
  n](t)$, and the AE potential evaluated on the self-consistent density $\Delta v\Hxc^{\rm AE}[\Delta
  n^{\rm AE}_{sc}](t)$. The exact $\Delta v\Hxc[\Delta n](t)$ is found by inserting the
exact density $\Delta n(t)$ into~\cite{LU08,FT12}
\begin{equation} \label{eq:vhxc}
  \Delta v\Hxc [\Delta n] = - \frac{\ddot{\Delta n} +4 T^2 \Delta n}{\sqrt{ 4T^2 \left(4 - \Delta n^2\right) - \dot{\Delta n^2}}} - \Delta v^0 -2 \mathcal{E}(t),
\end{equation}
This starts at its gs value $\Delta v\Hxc[\Delta n^0=1.9901,
  \dot{\Delta n}=\ddot{\Delta n}=0] \approx 1$ but soon increases
sharply and makes very large oscillations, which appear to be related
to maintaining non-interacting
$v$-representability~\cite{FT12,LU08,FM14,B08}: at the times of the
first sharp changes, the denominator in the first term of
Eq.~\ref{eq:vhxc} approaches zero, and the direction of the sharp
potential change is such to prevent the denominator actually becoming
zero.  Averaging through the oscillations, we see that the exact
$\Delta v\Hxc$ goes to $-\Delta v^0$ at $t \approx 224/U$, i.e. 
the exact $\Delta v\s$ becomes equal on the two sites (top panel). This is completely analogous with the
real-space case: there, a step in the HXC potential in the
intermolecular region develops such that when the CT state is reached,
the atomic levels of the donor(D) ion and acceptor(A) ion are ``re-aligned'' i.e. the step has size $\vert
I_D^{N_D-1}-I_A^{N_A+1}\vert$ in the large-separation limit~\cite{FERM13}.
In both real-space and Hubbard cases, it is the correlation potential
(lower left panel of Fig.~\ref{fig:vHxc_AE_AEsc}) that contains this
feature.  The oscillations in the exact $\Delta v\Hxc$ around its
average value near when the excited CT state is reached almost exactly
cancel the oscillations in the external field, as reflected in the
decreasing oscillations in the KS potential shown. This
field-counteracting effect is feature of the correlation potential and
appears also in the real-space case, related there to the absence of
polarization due to truncation to a few-level
system~\cite{EFRM12,LFSEM13}.

We now turn to the AE
calculations.  Consider first $\Delta v\Hxc^{\rm AE}[\Delta
  n](t)$.
In the real-space case of Ref.~\cite{FERM13}, as the CT state is
reached, the analogous AE correlation potential developed a step whose size, in the limit of large separation, approached
$\Delta_c^{D}(N-1)\equiv I_D^{N-1} - A_D^{N-1}$, the
derivative-discontinuity of the $(N-1)$-electron donor. The
 same occurs in the Hubbard model. First observe 
that $\Delta v\Hxc^{\rm AE}[\Delta n](t)$ shown in
Fig.~\ref{fig:vHxc_AE_AEsc} can be obtained by simply reading off the
potential from Fig.~\ref{fig:EHxc_vHxc_vc}, using the exact
instantaneous value of $\Delta n(t)$ of Fig~\ref{fig:dipoles} (left panel). 
So the shape of $\Delta v\Hxc^{\rm AE}[\Delta n](t)$ just tracks that of the $\Delta v\Hxc^{\rm gs} $ curve of
Fig~\ref{fig:EHxc_vHxc_vc}, moving from right to the center, gently oscillating around it.
Now, $\Delta n$ plays the role of the
density-variable as well as directly giving the particle number on
each site, $n_{\rm L,R} = 1 \pm \frac{\Delta n}{2}$. As a consequence, in the isolated-site
limit $T/U\to 0$, a
variation $\delta n$ near $\Delta n=0$ can be thought of as
adding(subtracting) a fraction of charge $\delta n$ to the one-fermion
site on the left(right):
\bea 
\label{eq:DD}
\nonumber &\left.2\frac{d E\c[\Delta
    n]}{d(\Delta n)}\right\vert_{\Delta n=0^+} - \left.2\frac{d
  E\c[\Delta n]}{d(\Delta n)}\right\vert_{\Delta n=0^-} =&\\ &\Delta
v\c^{\rm gs}[\Delta n = 0^+] - \Delta v\c^{\rm gs}[\Delta n = 0^-] \equiv 2
\Delta\c^{\rm 1-site}(N=1)\;& 
\eea 
where $E\c = E\Hxc - \frac{U}{8}(4 + \Delta n^2)$~\cite{FFTAKR13,CC13}.
The difference in the correlation potential on either side of $\Delta
n = 0$ therefore coincides with the derivative-discontinuity of the one
site with $(N-1)$ electrons; from Fig.~\ref{fig:EHxc_vHxc_vc}, this approaches the value
$\Delta\c^{\rm 1-site}(N=1) \approx 0.7$ for $T/U =0.05$. Returning to
Fig.~\ref{fig:vHxc_AE_AEsc}, as the CT state $\Delta n_{CT} \to 0$ is
approached, the exact $\Delta v\c[\Delta n]$ approaches $-\Delta v^0$,
while instead, the $\Delta v\c^{\rm AE}[\Delta n]$ tracks the
approaching discontinuity in Fig.~\ref{fig:EHxc_vHxc_vc}; as $T/U \to
0$, this change becomes sharper and larger, occuring over an ever
smaller region.  (The factor $2$ on the right of
Eq.~(\ref{eq:DD}) results from expressing the energy functional in
terms of the variable $\Delta n= n_L - n_R$, i.e. $\Delta v\c[\Delta
  n] = v\c^L[\Delta n] - v\c^R[\Delta n] =\frac{d E\c[\Delta
    n]}{d(\Delta n)} \frac{d \Delta n}{d n_L}-\frac{d E\c[\Delta
    n]}{d(\Delta n)} \frac{d \Delta n}{d n_R}$.)  So, in the limit, in
both the real-space molecule, and the Hubbard dimer, the AE
correlation potential in the CT state shifts the donor upwards
relative to the acceptor by an amount equal to the
derivative-discontinuity of the donor; in both cases, this
underestimates the shift provided by the exact correlation potential.

We now turn to $\Delta v\Hxc^{\rm AE}[\Delta n_{sc}](t)$ and $\Delta v\c^{\rm AE}[\Delta n_{sc}](t)$  in Fig.~\ref{fig:vHxc_AE_AEsc}.
Initially, $\Delta v\Hxc^{\rm AE}[\Delta n_{sc}](t)$ follows the exact
potential but very soon deviates from it: it makes small oscillations
near its initial value, hardly noticeable on the scale of the changes
of the exact $\Delta v\Hxc[\Delta n](t)$.
The dipole $\Delta n^{\rm AE}_{sc}$ is affected significantly only later (red dashed in left side Fig.~\ref{fig:dipoles}); the relatively large external potential seems to carry the dipole-oscillations with it for a while, before the
effect of the incorrect correlation potential is felt.
Certainly, the two sites never get anywhere close to being
re-aligned; {\it a stable CT state that has one electron on each therefore
cannot be approached}.

The failure to transfer the charge is not due to the error the AE
approximation makes for the CT excitation energy. In fact
$\omega^{AE}\approx \omega_{CT}$~\cite{FM14}. Reproducing accurately
the excitation spectrum is not enough for a functional to be able to
model time-resolved resonant CT dynamics.

\underline{\it os--os CT} 
The os--os  AE dipole fails miserably even after a
very short time, as shown on the right panel of Fig.~\ref{fig:dipoles}; 
one electron more or less always hovers on each site, while the exact propagation reaches the CT state at about $153/U$.
The KS, HXC and correlation 
potentials are shown on the right-hand-side of Fig.~\ref{fig:vHxc_AE_AEsc}. The initial
exact $\Delta v\c(t=0)$  exactly cancels the static external potential: $\Delta v\s(t=0)$ aligns the two sites, completely analogous to the
real-space case, where the correlation potential of a heteroatomic diatomic
molecule has a step that aligns the highest occupied
molecular orbital energies on each atom~\cite{P85b,GB96,TMM09}.  As charge
transfers, the exact $\Delta v\s(t)$ starts to oscillate on the optical scale, and there is a drop soon before the CT
excitation is reached. The  drop is related to the
denominator of $\Delta v\s(t)$ approaching zero as discussed before; in fact the shape resembles that of the cs--cs starting at the CT excited state.
The value of the exact $\Delta v\Hxc[\Delta n_{\rm CT}]$ can be obtained from
taking $\dot{\Delta n} = \ddot{\Delta n} = 0$ in Eq.~(\ref{eq:vhxc});
 note that it is  different from that obtained from 
its AE counterpart $\Delta v\Hxc^{\rm AE}[\Delta n = 1.9734]$ (middle right panel in Fig.~\ref{fig:vHxc_AE_AEsc}). This reflects the fact that the exact state is an excited state, not a gs of any potential.

The AE potential starts correctly, as it should for gs's, capturing
the alignment of the two sites, just as in the real-space case where
$v\c^{\rm AE}(\br)$ captures the initial intermolecular step. However,
$\Delta v\Hxc^{\rm AE}[n_{sc}](t)$ rapidly becomes a poor
approximation, hardly resembling the exact at all. The site-alignment creates a
near-degeneracy in the KS gs, unlike in the interacting system. The
true interacting gs though, has a Heitler-London form in the gs and
the CT excited state has a finite frequency.  This vanishing of the KS
gap implies that strong non-adiabaticity is required to open the gap
to the finite one of the interacting system~\cite{EGCM11}:
double-excitations are near-degenerate and critical to incorporate,
and non-adiabaticity is required. The nature of the states and
arguments above are the same as the real-space case, and so we expect that also for real molecules, a self-consistent AE
propagation will lead to a very poor dipole.  As for $\Delta
v\Hxc^{\rm AE}[\Delta n](t)$, it tracks $v\Hxc^{\rm gs}[\Delta n(t)]$
of Fig.~\ref{fig:EHxc_vHxc_vc} moving from near the center out to the
right; with  gentle oscillations reflecting the oscillations in
$\Delta n(t)$. Again we note that its value when the CT state is reached is the HXC potential of a
gs of density $\Delta n = 1.9734$ as opposed to the exact HXC
potential which is that for an excited-state of the same density.

A further similarity can be drawn between the real-space and Hubbard
models considering the static HXC kernel, $f\Hxc^{\rm gs}[\Delta n]
=d^2E\Hxc[\Delta n]/d(\Delta n)^2$ (Fig~\ref{fig:EHxc_vHxc_vc}). The
sharp peaked structure at $\Delta n = 0$ becomes proportional to a
$\delta$-function in the $T/U \to 0$ limit. The static kernel for real
os--os molecules at large separation~\cite{GGGB00,MT06} also diverges.
The exact non-adiabatic kernel $f\Hxc[\Delta n](\omega)$ must also diverge to open the gap, but there is a large non-adiabatic
correction to the static kernel in this case, and the AE 
frequencies are significantly different from those of the true
system~\cite{FM14}.

In both cases of CT, the form of the interacting state undergoes a
fundamental change: in the cs--cs case, from approximately a
single-Slater determinant initially to two determinants of
Heitler-London type in the CT state, while the reverse occurs for the
os--os case. The KS state however always remains a single determinant.
This gives the underlying reason for the development/loss of the
step structure in the exact potential in real-space, reflected in the
Hubbard model by the realignment of the two sites, signifying strong
correlation. An AE approximation captures this strong correlation
effect perfectly when it occurs in the gs (os--os case), but
our results shows it fails to propagate well even at short times due
to the near-degeneracy in the KS system.  In the cs--cs case, the AE
propagation begins accurately but ultimately fails to develop the
shift between donor and acceptor needed.

In summary, the asymmetric Hubbard dimer captures essential elements
of CT dynamics across a real-space molecule, enabling a decisive
verdict on the adiabatic approximation for time-resolved long-range CT
dynamics.  While previous work has shown the drastic
performance of usual adiabatic approximations~\cite{FERM13,Nest},
the present work shows that even propagating with the {\it best
  possible adiabatic} approximation, i.e.  adiabatically-exact,
fails. Accurately reproducing the CT frequency is not enough to model
fully time-resolved CT.  This suggests an urgent need to develop
non-adiabatic approximations for CT dynamics. The step feature in the
correlation potential, with non-local dependence on the density in
both space and time, must be modeled. There are obviously aspects of
CT in real molecules not captured in our model: the effect of many
electrons, three-dimensions, and coupling to ionic motion. These
 likely buffer the impact of the step, however there is no
reason to expect it will not still have significant consequences.

\begin{acknowledgments} 
JIF thanks Jordi Salvado and Mehdi Farzanehpour for useful conversations.
We gratefully acknowledge financial support from the National
Science Foundation CHE-1152784 (NTM) and US Department of Energy Office
of Basic Energy Sciences, Division of Chemical Sciences, Geosciences
and Biosciences under Award DE-SC0008623  (JIF).
\end{acknowledgments}

\end{document}